\documentclass[10pt,conference]{IEEEtran}

\usepackage{mathptmx}
\usepackage{mathtools}
\usepackage{graphicx}
\usepackage{algorithm}
\usepackage{algpseudocode}
\usepackage{amssymb}
\usepackage{amsmath}
\usepackage{color}
\usepackage{array}
\usepackage{tabularx}
\usepackage{float}
\usepackage{multirow}
\usepackage{subfigure}
\usepackage{balance}
\usepackage{epsfig}
\usepackage{epstopdf}
\usepackage{etoolbox}
\usepackage[strings]{underscore}
\usepackage[labelsep=period]{caption}

\pdfoutput=1

\newcommand\blfootnote[1]{%
  \begingroup
  \renewcommand\thefootnote{}\footnote{#1}%
  \addtocounter{footnote}{-1}%
  \endgroup
}

\begin{document}
\title{A comparative analysis of state-of-the-art SQL-on-Hadoop systems for interactive analytics}
\author{\IEEEauthorblockN{Ashish Tapdiya}
\IEEEauthorblockA{Vanderbilt University\\
Email: ashish.tapdiya@vanderbilt.edu}
\and
\IEEEauthorblockN{Daniel Fabbri}
\IEEEauthorblockA{Vanderbilt University\\
Email: daniel.fabbri@vanderbilt.edu}
}

\maketitle

\thispagestyle{plain}
\pagestyle{plain}

\begin{abstract}
Hadoop is emerging as the primary data hub in enterprises, and SQL represents the de facto language for data analysis. This combination has led to the development of a variety of SQL-on-Hadoop systems in use today. While the various SQL-on-Hadoop systems target the same class of analytical workloads, their different architectures, design decisions and implementations impact query performance. In this work, we perform a comparative analysis of four state-of-the-art SQL-on-Hadoop systems (Impala, Drill, Spark SQL and Phoenix) using the Web Data Analytics micro benchmark and the TPC-H benchmark on the Amazon EC2 cloud platform. The TPC-H experiment results show that, although Impala outperforms other systems (4.41x -- 6.65x) in the text format, trade-offs exists in the parquet format, with each system performing best on subsets of queries. A comprehensive analysis of execution profiles expands upon the performance results to provide insights into performance variations, performance bottlenecks and query execution characteristics.
\end{abstract}
\section{Introduction}
\blfootnote{\hspace{-8 pt}A short version of this paper appeared in IEEE Big Data Conference 2017. \\ Available Online: http://ieeexplore.ieee.org/abstract/document/8258066/}
Enterprises are increasingly using Hadoop as a central repository to store the data generated from a variety of sources including operational systems, sensors, social media, etc. Although general purpose computing frameworks such as MapReduce (MR) \cite{mr} and Spark \cite{spark-paper} enable users to perform arbitrary data analysis on Hadoop, users remain comfortable with and rely on SQL to glean actionable insights from the data. This combination has led to the development of several SQL-on-Hadoop systems, each of which have their own architectures, design decisions and implementations.

SQL-on-Hadoop systems take various forms. One class of system relies on a batch processing runtime for query execution. Systems like Shark \cite{shark} and Spark SQL \cite{sparksql} employ the Spark runtime to execute queries specified using the standard SQL syntax. Hive \cite{hive} enables users to write queries in the HiveQL language and compiles it into a directed acyclical graph (DAG) of jobs that can be executed using MR or Spark or Tez \cite{tez} runtime. Another class of SQL-on-Hadoop system is inspired by Google's Dremel \cite{dremel}, and leverages a massively parallel processing (MPP) database architecture. Systems like Impala \cite{impala} and Drill \cite{drill-paper} avoid the overhead associated with launching jobs for each query by utilizing long running daemons. However, even within this class, differing design decisions, such as early versus late materialization, impact query performance.

In this work we evaluate and compare the performance of four state of the art SQL-on-Hadoop query engines for interactive analytics: Impala, Drill, Spark SQL and Phoenix \cite{Phoenix}. We chose to study these systems due to a variety of reasons: 1) Each evaluated system has a large user base as they are part of major Hadoop distributions including Cloudera, MapR, Hortonworks etc. 2)  Each system is open source, targets the same class of interactive analytics and is optimized for the same storage substrate. 3)  Each system employs cost based optimization and advanced run time code generation techniques. 4) These systems have significantly different architectures (ex. batch processing versus long running daemons) and make varying design decisions for query processing (ex. vectorized \cite{vector} versus volcano model \cite{volcano}).

Our goal is to evaluate and understand the characteristics of two primary components of a SQL-on-Hadoop system (query optimizer and query execution engine) and their impact on the query performance. For the query optimizer, our objective is to characterize the execution plan generation, the join order selection and the operator selection in each evaluated system. For the query execution engine, we aim to evaluate the efficiency of operator implementations in each system and identify the performance bottlenecks by examining query execution profiles. We design experiments to understand specific characteristics of each system: scale-up, size-up, and the impact of file formats (text versus parquet).

We use the Web Data Analytics (WDA) micro benchmark \cite{pavlo} and the TPC-H benchmark \cite{tpch} to experimentally evaluate systems. We select the Amazon EC2 cloud platform as our experiment environment so that the results from future studies that benchmark new SQL-on-Hadoop systems can be compared with our results. Our experiment results show that,
\begin{itemize}
\item Drill exhibits the lowest join and aggregation operator times across all evaluated systems. In Drill, the scan operator contributes the most to the query response time (RT) in the parquet storage format and becomes a performance bottleneck in the text storage format.
\item Phoenix is well suited for data exploration tasks (such as selection and aggregation) and gains notable performance boost through range-scans. However, the client coordinated data-exchange operation is the principal performance bottleneck in Phoenix, making it ill-suited for join heavy workloads that shuffle large amounts of data.
\item Impala has the most efficient and stable disk I/O subsystem among all evaluated systems; however, inefficient CPU resource utilization results in relatively higher processing times for the join and aggregation operators. 
\item The scan and join operators are the chief contributors to the query RT in Spark SQL. In addition, garbage collection (GC) time represents a notable fraction of the query RT.
\item \textit{\textsc{scale-up.}} Impala exhibits linear scale-up behavior. Drill and Phoenix show super-linear scale-up behavior due to an increase in the scan and the data-exchange operator times, respectively. Increased join operator time results in marginally super-linear scale-up in Spark SQL.     
\item \textit{\textsc{size-up.}} A sub-linear increase in the scan operator time leads to sub-linear size-up in Impala. Drill shows linear size-up for larger database sizes. Spark SQL shows sub-linear sizeup behavior due to the sub-linear increase in all operator times except the data-exchange operator.  
\end{itemize}
%
\begin{table*}[ht]
\centering
\caption{Qualitative comparison of evaluated SQL-on-Hadoop systems.}
\label{qual}
\scalebox{0.7}{
\begin{tabular}{|c|c|c|c|c|c|}
\hline
\textbf{System}  & \textbf{Query Optimizer}                                                                                       & \textbf{Execution Model}                                                                                                          & \textbf{Fault Tolerance}                                                    & \textbf{Execution Runtime}                                                               & \textbf{Schema Requirements}                                                                  \\ \hline
\textbf{Impala}  & \begin{tabular}[c]{@{}c@{}}Cost Based Optimizer that attempts \\ to minimize network transfer\end{tabular}    & \begin{tabular}[c]{@{}c@{}}Volcano Batch-at-a-Time, Pipelined \\ Execution, Runtime Code Generation\end{tabular}                  & Requires Query Restart                                                      & \begin{tabular}[c]{@{}c@{}}MPP Engine with\\  Long Running Daemons\end{tabular}          & Upfront definition required                                                                   \\ \hline
\textbf{Spark SQL}   & \begin{tabular}[c]{@{}c@{}}Extensible Catalyst Optimizer with \\ Cost and Rule based optimization\end{tabular} & \begin{tabular}[c]{@{}c@{}} Pipelined Execution, Whole \\ Stage Code Generation, Vectorized Processing\end{tabular} & \begin{tabular}[c]{@{}c@{}} Lineage based\\RDD Transformations \end{tabular} & Spark Engine                                                                             & \begin{tabular}[c]{@{}c@{}}Reflection and Case classes\\  based schema inference\end{tabular} \\ \hline
\textbf{Drill}   & \begin{tabular}[c]{@{}c@{}}Apache Calcite derived Cost \\ Based Optimizer\end{tabular}                         & \begin{tabular}[c]{@{}c@{}}Vectorized Processing, Pipelined Execution,\\  Runtime Code Compilation\end{tabular}                   & Requires Query Restart                                                      & \begin{tabular}[c]{@{}c@{}}MPP Engine with\\  Long Running Daemons\end{tabular}          & Schemaless Query Support                                                                      \\ \hline
\textbf{P-HBase} & \begin{tabular}[c]{@{}c@{}}Apache Calcite derived Cost \\ Based Optimizer\end{tabular}                         & Blocking Execution                                                                                                                & Requires Query Restart                                                      & \begin{tabular}[c]{@{}c@{}}Client Coordinated Parallel \\ HBase Scans Based\end{tabular} & Upfront definition required                                                                   \\ \hline
\end{tabular}
}
\end{table*}
\section{Background}
\label{sec:backg}
In this section we review the SQL-on-Hadoop systems evaluated in this study. Table \ref{qual} presents a qualitative comparison of the evaluated systems.
\subsection{Evaluated Systems}
\label{sec:evalsys}
\textbf{\textsc{impala.}} Impala is a MPP query execution engine inspired by the Dremel system. The Impala runtime comprises of long running daemons that are generally co-located with the HDFS data nodes. Impala makes extensive use of LLVM library to gain CPU efficiency by reducing the performance impact of virtual function calls and generating query specific code at runtime for functions that are called numerous times for a single query (ex. record parser). 
The Impala execution engine harnesses a volcano model with fully pipelined execution.

\textbf{\textsc{spark sql.}} Spark SQL is a component in the Spark ecosystem that is optimized for structured data processing. DataFrames represent the primary abstraction in Spark SQL. A DataFrame is a distributed collection of data organized into named columns and is analogous to a table in the relational database (DB). The Tungsten query execution engine in Spark SQL achieves CPU efficiency and bare metal processing speed through whole-stage code generation; however, if the whole-stage code generation is not possible (for ex. for third party code) then the Tungsten engine harnesses vectorized processing to exploit the SIMD support in modern CPUs. 

\textbf{\textsc{drill.}} Drill is a MPP query execution engine inspired by the Dremel system. Drill optimizes for columnar storage as well as columnar execution through an in-memory hierarchical columnar data model. The Drill execution engine utilizes vectorized query processing to achieve peak efficiency by keeping CPU pipelines full at all times. The run time code compilation enables Drill to generate efficient custom code for each query. Drill has the ability to discover schema on the fly.

\textbf{\textsc{phoenix.}} Phoenix is a SQL skin on top of HBase \cite{HBase}. The client embedded JDBC driver in Phoenix transforms the SQL query into a series of HBase scans and coordinates the execution of scans to generate result-set (RS). The Phoenix execution engine harnesses HBase features such as scan predicate pushdown and coprocessors to push processing on the server side. We use P-HBase to refer to the system resulting from the combination of Phoenix and HBase systems. 

For an aggregation query, the Phoenix client issues parallel scans to the HBase region servers. The results of each scan are partially-aggregated on the server using an aggregation coprocessor. The partially-aggregated results are then merged in the client to produce the final RS. For a join query, 1) client issues parallel scans to read one input of the join operation and prepare the hash table, 2) the prepared hash table is then sent to and cached on each region server hosting regions of the other input, 3) the client then scans the other input of the join operator, 4) the scanned records are then joined with the matching records in the cached hash table on each region server using the join coprocessor, and 5) the joined records from each region server are then merged in the client.
\subsection{Profiling Tools}
\label{sec:profiling}
In this section we elucidate on how we utilize the profiling information exposed by each evaluated system.

\textbf{\textsc{impala}} and \textbf{\textsc{drill.}} The profiler provides an execution summary for the scan, join, aggregation, data-exchange and sort operators present in a query execution plan. Note, the scan operator includes the time to scan, filter, and project tuples from a table. Also, the data-exchange operator includes the time to transfer the data over the network; however, the data de/serialization time is not summarized by the profiler.

\textbf{\textsc{spark sql.}} The profiler generates detailed statistics for each execution stage in the query DAG. For each stage, we summarize the task data to calculate average values for scheduling delay, GC time, shuffle-read time, shuffle-write time and executor-computing time. We map the query execution plan operators to the query DAG stages. Note, multiple operators (with pipelined execution), such as join and partial-aggregation, final-aggregation and sort, scan and partial-aggregation, may be mapped to a single DAG stage. The executor-computing time for a stage is credited as the processing time for the operator/s mapped to that stage. The Spark runtime performs I/O in the background while a task is computing, and shuffle-read represents the time for which a task is blocked reading the data over the network from another node \cite{kay}. Hence, the shuffle-read time for a stage is attributed as the processing time for the corresponding data-exchange operator/s in the query plan. Also, the shuffle-write time for a stage is assigned to the data serialization overhead.

In this study, we utilize the average operator time in each evaluated system for analysis. In addition, we could not use operator level execution time break-down in \textbf{\textsc{phoenix}} since currently it does not record execution statistics. 

We acknowledge the variance in profiling information exposed by evaluated systems; however, despite the differences, we are able to gain significant insight into the performance characteristics of evaluated systems.
\begin{figure}[h]
\includegraphics[clip=true, trim=5 278 340 23, scale=.3]{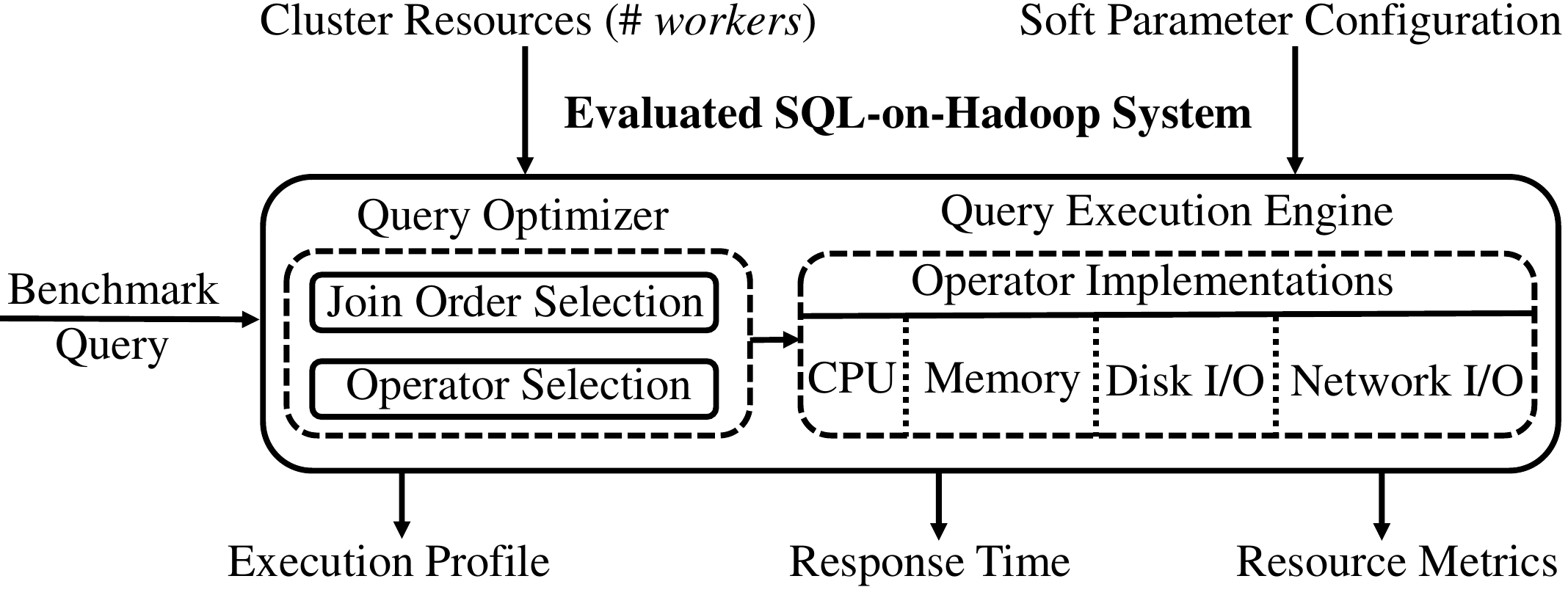}
\centering
\caption{\small Query execution model in each evaluated system.}
\label{fig:model}
\end{figure}
\section{Experiment Goals}
In this section we describe our study goals. The query performance in a SQL-on-Hadoop system is dependent both on the quality of the execution plan generated by the query optimizer and the efficient execution of the generated plan by the query engine, as shown in Figure \ref{fig:model}. The optimizer generates a query execution plan by evaluating different join orders and selecting physical operators for the relational algebra operations in the chosen query plan. The query execution engine utilizes operator implementations to carry out the generated plan and produce the output RS. An operator may use one or more sub-systems (CPU, Disk, Network and Memory) in the execution engine to perform its task. The query RT represents our performance metric. We collect resource utilization metrics and query execution profiles in each system.

In this study, our goal is to evaluate and understand the characteristics of two main components of a SQL-on-Hadoop system (query optimizer and query execution engine) and their impact on the query performance. For the query optimizer, our objective is to characterize the execution plan generation, the join order selection and the operator selection in each evaluated system. To this end, we analyze and compare the generated execution plans in each system and understand their impact on the query performance. For the query execution engine, we aim to evaluate the efficiency of operator implementations in each system and identify the performance bottlenecks. To this end, we utilize the query execution profiles to extract the operator processing times. In each system, we aggregate the processing times for each operator type to understand the contribution of each operator type to the query RT. In addition, we compare the total benchmark operator processing times between evaluated systems to identify if a sub-system in an execution engine is a performance bottleneck. 
 
We perform experiments along three dimensions -- \textit{storage format, scale-up} and \textit{size-up}. We examine the impact of text (row-wise) and parquet (columnar) storage formats on the query performance in each system. To understand the \textit{size-up} characteristic in each system, we increase the data size in a cluster and examine the query performance changes. We evaluate the \textit{scale-up} behavior in each system by proportionally increasing both the cluster and the data size. Note that we use the \textit{scale-up} and \textit{size-up} definitions specified in \cite{djdewitt}.
\section{Micro Benchmark Experiments}
In this section we utilize the WDA micro benchmark proposed in \cite{pavlo} to evaluate and compare the performance of Impala, Drill, Spark SQL and P-HBase systems.  
\subsection{Hardware Configuration} 
\label{sec:hw-conf}
We harness Amazon EC2 cloud environment to setup experiment testbeds. Our testbeds comprise of ``\textit{worker}'' VMs of r3.4xlarge instance type, a ``\textit{client}'' VM of  r3.4xlarge instance type and a ``\textit{master}'' VM of d2.xlarge instance type. A r3.4xlarge VM instance is configured with 16 vCPUs, 122GB RAM, 320GB SSD instance storage and 120GB EBS storage and a d2.xlarge VM instance is configured with 4 vCPUs, 30.5GB RAM, 3x 2000GB HDD instance storage and 120GB EBS storage. Note, we choose r3.4xlarge instance for the \textit{worker} nodes since evaluated systems are optimized for in-memory execution and the RAM size in a r3.4xlarge instance is large enough to hold the intermediate result-sets of all workload queries in-memory for the largest evaluated DB size.

\subsection{Software Configuration}
\label{sec:sw-conf}
We deploy HDFS, HBase, Drill, Impala, Spark, Yarn, Hive and Zookeeper frameworks on each cluster. We host the control processes (ex. HMaster, NameNode etc.) from each framework on the \textit{master} VM. We also use \textit{master} VM as the landing node for the data generation. Worker processes from each framework (ex. DataNode, Drillbit etc.) are hosted on each \textit{worker} VM in the cluster. We reserve the \textit{client} VM to drive the workload. We deploy the Phoenix JDBC driver in the \textit{client} VM and put relevant jars on the classpath of each Region Server. We use the Cloudera Distribution v5.8.0, Impala v2.6.0, Spark v2.0, Drill v1.8.0, HBase v1.2.0 and Phoenix v4.8.0. 

Fine tuning of soft parameters is quintessential to getting the best performance out of a software system. To identify the values for performance knobs that ensure a fair comparison between evaluated systems, we rely on: 1) Settings used in prior studies (ex. \cite{icpe, floratou2014}). 2) Best practice guidelines from industry experts and system developers (ex. \cite{tuning}). 3) Experimentation with different values that enable us to run all benchmark queries and achieve the best performance. 

For HDFS, we enable short-circuit reads, set the block size to 256MB and set the replication to 3. We enable Yarn as the resource manager for Spark and set \textit{spark.executor.cores} to 8, \textit{spark.executor.memory} to 8GB and \textit{spark.memory.offheap} to 16GB. We set the heap size to 20GB in each Region-Server. We assign 95GB to each worker process in Impala and Drill.
\subsection{Experiment Setup}
\label{sec:ex-setup}
We evaluate systems one at a time. During the evaluation of a system, we stop the processes for other systems to prevent any interference. We disable HDFS caching in each system. We execute benchmark queries using a closed workload model, where a new query is issued after receiving the response for the previous query. We run each query five times and report the average of the query RT for the last four runs with a warm OS cache. We use a JDBC driver to run the workload in P-HBase and shell scripts to issue queries in Spark SQL, Drill and Impala. Similar to \cite{pavlo}, we write query output to HDFS in Spark SQL and to local file system in Impala, Drill and P-HBase. We use \textit{collectl} linux utility to record resource utilization on each \textit{worker} node with a 3 second interval. We export the query execution profile in all systems except P-HBase since it currently does not record execution statistics.
\subsection{WDA Benchmark}
The benchmark schema comprises of two relations (UserVisits and Rankings) that model the log files of HTTP server traffic. The benchmark workload comprises of simple tasks (selection query, aggregation query, and join query) related to HTML document processing. Note, similar to \cite{amp}, we choose the aggregation task variant that groups records by the source IP address prefix. In addition, we remove the UDF task from the workload since it could not be implemented in each evaluated system. We use the data generator utility provided with the WDA benchmark to generate 20GB of UserVisits data and 1GB of Rankings data per \textit{worker} node (same as in \cite{pavlo}). We refer reader to \cite{pavlo} for detailed benchmark description.
\begin{table}[h]
\centering
\vspace*{1mm}
\caption{Data preparation times (seconds) in evaluated systems for WDA benchmark.}
\label{load-wdab}
\scalebox{.65}{
\begin{tabular}{|c|c|c|c|c|c|c|c|c|}
\hline
\multirow{2}{*}{\textbf{\begin{tabular}[c]{@{}c@{}}\# of \\ Worker \\ Nodes\end{tabular}}} & \multirow{2}{*}{\textbf{\begin{tabular}[c]{@{}c@{}} \\ Insert \\ HDFS\end{tabular}}} & \multicolumn{3}{c|}{\textbf{Impala}}                                                                                                                 & \multicolumn{3}{c|}{\textbf{P-HBase}}                                                                                                         & \textbf{\begin{tabular}[c]{@{}c@{}}Drill and \\ Spark SQL\end{tabular}} \\ \cline{3-9} 
                                                                                           &                                                                                  & \textbf{\begin{tabular}[c]{@{}c@{}}Load \\ Tables\end{tabular}} & \textbf{\begin{tabular}[c]{@{}c@{}}Compute \\ Stats\end{tabular}} & \textbf{Total} & \textbf{\begin{tabular}[c]{@{}c@{}}Load \\ Tables\end{tabular}} & \textbf{\begin{tabular}[c]{@{}c@{}}Major\\ Compact\end{tabular}} & \textbf{Total} & \textbf{Total}                                                          \\ \hline
\textbf{2}                                                                                 &  280                                                                &         10.93                                                        &                                   283.97          &   574.9             &                                     6748                            &      1590                                                            &   8618             &                280                                                         \\ \hline
\textbf{4}                                   &            550                                                   &       9.5                                                          &                     291.4                                              &     850.9           &       7027                                                          &                   1722                                               &        9299        &         550                                                                \\ \hline
\textbf{8}                                          &         1124                                                              &            8.6                                  &               295.2                                   &       1427.8         &                      9088                                  &                      1559                                            &        11771        &               1124                                                          \\ \hline
\end{tabular}
}
\end{table}
\vspace*{-3mm}
\subsection{Data Preparation}
\label{wdab-prep}
We evaluate each system with the data stored in the text format to ensure storage format parity across systems. In each system, we first load the text data from the landing node into HDFS. Drill and Spark SQL are capable of directly querying the text data stored in HDFS. Next, we describe the subsequent data preparation steps taken in Impala and P-HBase,

\textbf{\textsc{impala.}} We create the benchmark schema and load tables with the text data stored in HDFS. Next, we utilize the \textsc{compute stats} command to collect statistics for each table.

\textbf{\textsc{p-hbase.}} We create UserVisits and Rankings tables with visitDate and pageRank as the first attribute in the respective row-keys. As a result, similar to \cite{pavlo}, UserVisits and Rankings tables are sorted on visitDate and pageRank columns respectively. We assign all columns in a table to a single column family. We utilize the \textit{salting} feature provided by Phoenix to pre-split each table with two regions per region-server. \textit{Salting} prevents region-server hotspotting and achieves uniform load distribution across region-servers in the cluster. We utilize a MR based bulk loader in Phoenix to load the text data stored in HDFS into HBase tables. Next, we run \textit{major compaction} on each table to merge multiple HFiles of a region into a single HFile. Phoenix collects data statistics for each table during the major compaction process.

Table \ref{load-wdab} presents the data preparation times in each evaluated system for 2, 4 and 8 \textit{worker} nodes. The time to load data into HDFS from the landing node increases linearly with an increase in the cluster and the data size (recall that the data size increases with each \textit{worker}). The time to load data into Impala tables is a small fraction of the total data preparation cost since Impala's text loading process simply moves the files to a Impala managed directory in HDFS. The statistics computation process in Impala and the major compaction process in P-HBase exhibit good \textit{scale-up} characteristic as the execution times remain nearly constant for the different cluster sizes. However, the data loading times for the MR based bulk loader in P-HBase increase with an increase in the data size and the number of \textit{worker} nodes in the cluster.
\begin{table*}[th]
\centering
\caption{Query RTs (in seconds) in evaluated systems using WDA benchmark for 2, 4 and 8 worker nodes in the cluster. RT in bold text denotes the fastest system for each query and cluster size combination. AM denotes the arithmetic mean. To compute the normalized AM: for each query, we normalize the query RTs in each system and for each cluster size by the query RT in Impala with 2 worker nodes.}
\label{pavlo-numbers}
\scalebox{.7}{
\begin{tabular}{|c|c|c|c|c|c|c|c|c|c|c|c|c|c|}
\hline
\multicolumn{2}{|c|}{\multirow{2}{*}{\textbf{WDA Query No.}}} & \multicolumn{4}{c|}{\textbf{\# of worker nodes -- 2}}                       & \multicolumn{4}{c|}{\textbf{\# of worker nodes -- 4}}                       & \multicolumn{4}{c|}{\textbf{\# of worker nodes -- 8}}                       \\ \cline{3-14} 
\multicolumn{2}{|c|}{}                   & \textbf{Impala} & \textbf{Spark} & \textbf{Drill} & \textbf{P-HBase}  & \textbf{Impala} & \textbf{Spark} & \textbf{Drill} & \textbf{P-HBase}  & \textbf{Impala} & \textbf{Spark} & \textbf{Drill} & \textbf{P-HBase}  \\ \hline
\multirow{3}{*}{\textbf{Q1}}             & Execution           & \textbf{0.28}            & 5.5              & 9              & 0.3                     & \textbf{0.28}           & 6            & 16.5           & 0.34                    & \textbf{0.28}            & 7              & 30             &         0.62                \\ \cline{2-14} 
                                         & Write Output        & 3.13            & \textbf{0.5}            & 2              & 8.7                     & 6.03            & \textbf{0.9}            & 3.5            & 14.66                   & 11.52           & \textbf{2.1}            & 5              &        26.38                 \\ \cline{2-14} 
                                         & Total               & \textbf{3.41}            & 6            & 11             & 9                       & \textbf{6.31}            & 6.9            & 20             & 15                      & 11.8            & \textbf{9.1}           & 35             &   27                         \\ \hline
\textbf{Q2}                         & Total               & \textbf{12.6}            & 68.6           & 33             & 240                     & \textbf{12.4}            & 68             & 31.8           & 228                     & \textbf{13.15}           & 69.1           & 68             &            231             \\ \hline
\textbf{Q3}                              & Total               & 14.8            & 67.1           & 34             & \textbf{14}                      & \textbf{15}              & 70.1           & 35.1           & 17                      & \textbf{15.12}           & 84.6           & 72             &           40              \\ \hline
\multicolumn{2}{|c|}{\textbf{AM}}      & \multicolumn{1}{c|}{10.27}  & \multicolumn{1}{c|}{47.23} & \multicolumn{1}{c|}{26}    & \multicolumn{1}{c|}{87.67} & \multicolumn{1}{c|}{11.24}  & \multicolumn{1}{c|}{48.33} & \multicolumn{1}{c|}{28.97} & \multicolumn{1}{c|}{86.67} & \multicolumn{1}{c|}{13.36}  & \multicolumn{1}{c|}{54.27} & \multicolumn{1}{c|}{58.33} & \multicolumn{1}{c|}{99.33} \\ \hline
\multicolumn{2}{|c|}{\textbf{Normalized AM}}      & \multicolumn{1}{c|}{1}  & \multicolumn{1}{c|}{3.89} & \multicolumn{1}{c|}{2.7}    & \multicolumn{1}{c|}{7.52} & \multicolumn{1}{c|}{1.28}  & \multicolumn{1}{c|}{4.04} & \multicolumn{1}{c|}{3.58} & \multicolumn{1}{c|}{7.84} & \multicolumn{1}{c|}{1.84}  & \multicolumn{1}{c|}{4.61} & \multicolumn{1}{c|}{6.79} & \multicolumn{1}{c|}{9.63} \\ \hline
\end{tabular}
}
\end{table*}
\begin{figure*}[t]\centering
\hspace*{-2.0 in}
\begin{minipage}[t]{0.24\textwidth}\centering
\subfigure{
\includegraphics[clip=true, trim=20 175 30 14, scale=.33]{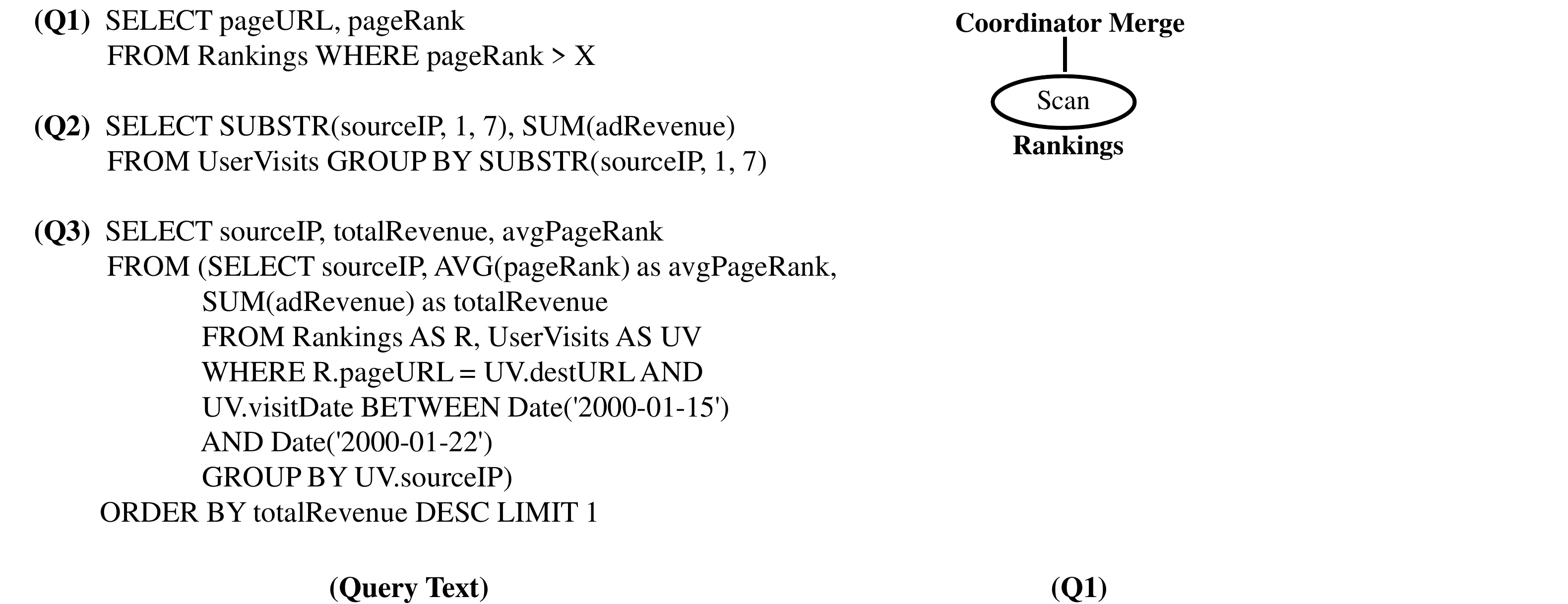} %
\label{fig:q4-1}
}
\end{minipage}
\hspace*{1.7 in}
\begin{minipage}[t]{0.24\textwidth}\centering
\subfigure{
\includegraphics[clip=true, trim=30 175 40 14, scale=0.33]{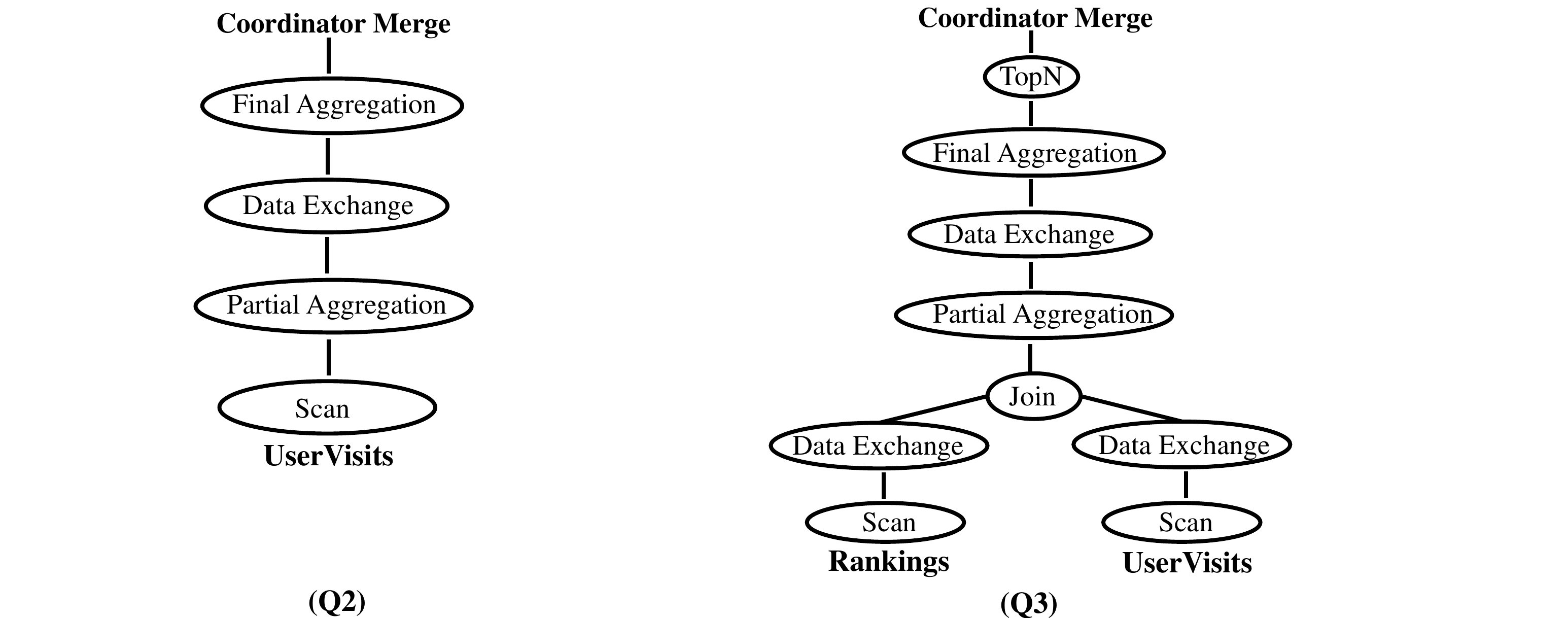}
\label{fig:q4-2}
}
\end{minipage}
\caption{\small Query text and logical plans for the WDA benchmark queries.}
\label{fig:logical}
\end{figure*}
\subsection{Experiment Results}
Table \ref{pavlo-numbers} presents the query RT for tasks in the WDA benchmark. Note, the standard error of the mean query RT is negligible in evaluated systems; hence, we exclude it from the presentation of the results. To evaluate the \textit{scale-up} characteristic in each system, we perform experiments with 2, 4 and 8 \textit{worker} nodes in the cluster. On an average, Impala is (5.2x - 7.5x), (2.7x - 3.7x) and (2.5x - 3.9x) faster as compared to P-HBase, Drill, and Spark SQL, respectively. Next, for each benchmark task, we analyze the execution profiles to understand its performance characteristics in each evaluated system. Figure \ref{fig:logical} shows the text and the logical execution plans for the benchmark tasks.
\subsubsection{Selection Task (Q1)}
The selection task is a scan query with a lightweight filter on the Rankings table. It is designed to primarily measure the read and the write throughput in each evaluated system. Similar to \cite{pavlo}, we set the filter parameter to 10, resulting in $\approx$ $33K$ records per \textit{worker} node in the cluster.

\textbf{\textsc{discussion.}} Impala, Spark SQL and Drill perform a full scan of the Rankings table and apply the filter to generate the output RS. Impala is the fastest with sub-second scan operator time. P-HBase achieves significant scan efficiency by performing a range-scan on the Rankings table since it is sorted on the pageRank attribute. The scan operator time represents a major fraction (.95x) of the total query execution time in Spark SQL. Drill is the slowest and the scan operator is the primary contributor (.9x) to the total query execution time in Drill. Impala and Spark SQL exhibit the lowest (8\%) and the highest (35\%) mean CPU utilizations, respectively. The output RS materialization time shows that Spark SQL achieves the highest write throughput and P-HBase is the slowest.

\textbf{\textsc{scale-up behavior.}} The constant query execution time across different cluster sizes shows linear \textit{scale-up} behavior in Impala. Although the relative query execution time in Drill, Spark SQL and P-HBase increases with the increase in the cluster size, Drill exhibits the maximum increase ($\approx$ 80\%). Further examination of the query execution profile in Drill shows that, although the processing time of the scan operator remains nearly constant, the scan operator wait time increases as the cluster is scaled up, resulting in the observed increase in the query execution time. In each system, the output RS materialization time increases proportionately with the increase in the cluster size since nearly $33K$ records are generated for each \textit{worker} node in the cluster.  
\subsubsection{Aggregation Task (Q2)}
The aggregation task groups records by the seven-character prefix of the sourceIP attribute in the UserVisits table and produces $\approx$ 1000 groups regardless of the number of  \textit{workers} in the cluster. It is designed to evaluate the performance of each system for parallel analytics on a single table. Note, output RS materialization time is negligible; hence, we only report the total query RT.

\textbf{\textsc{discussion.}} Each evaluated system scans the UserVisits table and performs partial-aggregation followed by the final-aggregation to generate the output RS. Impala is at least 5x faster than the other evaluated systems. Impala uses the streaming-aggregation (SA) operator and it is the primary contributor ($\approx$ .85x) to the query RT. Drill utilizes the hash-aggregation (HA) operator and although the query RT is high in Drill, the total aggregation operator processing time is less than 2s ( as compared to $\approx$ 11s in Impala) across all cluster sizes. The scan operator is the primary contributor (.65x - .9x) to the query RT in Drill. P-HBase exhibits the highest query RT since it scans approximately 40\% more data (UserVisits size is $\approx$ 1.4x in P-HBase, due to the HBase HFile format that appends key, column family, column and timestamp to each cell value) than other systems and its execution engine lacks the run time code generation feature. In Spark SQL, the scan and the partial-aggregation operators are mapped to a single DAG stage that contributes nearly 98\% to the query RT. Impala achieves lowest mean CPU utilization (20\%) as compared to Drill (35\%), Spark SQL (60\%) and P-HBase (55\%).

\textbf{\textsc{scale-up behavior.}} Impala, Spark SQL and P-HBase exhibit near linear \textit{scale-up} characteristic as the query RT remains almost constant with the increase in the cluster size. On the contrary, the query RT in Drill more than doubles as the cluster size is increased from 4 to 8 \textit{worker} nodes. Further analysis of execution profile shows that, similar to the selection task, increases in the scan operator wait time is primarily responsible for this increase in the query RT. Note, although the query RT in Drill increases as the cluster is scaled up, the aggregation operator time remains nearly constant.
\subsubsection{Join Task (Q3)}
The join task consumes two input tables (Rankings and UserVisits) and combines records with values matching on the join attributes (pageURL and destURL). It is designed to examine the efficiency of each sub-system (CPU, disk, etc.) in the evaluated query execution engines.

\textbf{\textsc{discussion.}} Drill, Impala, and Spark SQL scan and hash-partition Rankings and UserVisits tables on the join keys. The matching records from the partitioned tables are then joined, aggregated and sorted to generate the output RS. Impala utilizes the hash-join (HJ) operator and although the join operator processing time is high in Impala ($\approx$ 7s), query RT is dominated by the scan operator time ($\approx$ 80\% of query RT) for the UserVisits table. Similarly, the scan operator time for the UserVisits table is the primary contributor to the query RT in both Spark SQL (.8x -- .9x) and Drill (at least .75x). Drill uses the HJ operator and despite the high query RT, Drill exhibits the lowest join operator processing time (less than 4.5s) among all evaluated systems and across all cluster sizes. P-HBase performs a range-scan of the UserVisits table to prepare and broadcast the hash table to each region server. Since the UserVisits table is sorted on the filter attribute visitDate, P-HBase is able to perform the range-scan and gain significant scan efficiency.

\textbf{\textsc{scale-up behavior.}} The query RT in P-HBase increases as the cluster is scaled up since the time to broadcast the hash table of one join input from the client to the region servers increases. Spark SQL utilizes the sort-merge-join (SMJ) operator and the join operator processing time shows increase as the cluster is scaled up. Similar to the aggregation task, the query RT in Drill more than doubles as the cluster size is increased from 4 to 8 \textit{workers} due to an increase in the scan operator wait time. In addition, the join operator time in Drill shows marginal increase (2.7s -- 4.2s) as the cluster is scaled up. Impala exhibits near linear \textit{scale-up} behavior with almost constant query RTs across different cluster sizes.  
\begin{table*}[ht]
\centering
\caption{Query RTs (seconds) in evaluated systems using the TPC-H benchmark at 125, 250 and 500 scale factors. RT in bold text denotes the fastest system for each query, scale factor and file format combination. To compute the normalized AM--Q\{2,11,13,16,19,21,22\}: for each query, we normalize the query RTs in each system, at all scale factors and for each storage format by the query RT in Impala, for the parquet storage format, at scale factor 125.}
\label{tpch-numbers}
\scalebox{0.685}
{
\hspace*{-0.08 in}
\begin{tabular}{|c|c|c|c|c|c|c|c|c|c|c|c|c|c|c|c|c|c|c|}
\hline
\multirow{3}{*}{\textbf{\begin{tabular}[c]{@{}c@{}}TPC-H\\  Query\\ No.\end{tabular}}} & \multicolumn{6}{c|}{\textbf{TPC--H Scale Factor -- 125}}                                                             & \multicolumn{6}{c|}{\textbf{TPC--H Scale Factor -- 250}}                                                             & \multicolumn{6}{c|}{\textbf{TPC--H Scale Factor -- 500}}                                                             \\ \cline{2-19} 
                                                                                       & \multicolumn{2}{c|}{\textbf{Impala}} & \multicolumn{2}{c|}{\textbf{Spark SQL}} & \multicolumn{2}{c|}{\textbf{Drill}} & \multicolumn{2}{c|}{\textbf{Impala}} & \multicolumn{2}{c|}{\textbf{Spark SQL}} & \multicolumn{2}{c|}{\textbf{Drill}} & \multicolumn{2}{c|}{\textbf{Impala}} & \multicolumn{2}{c|}{\textbf{Spark SQL}} & \multicolumn{2}{c|}{\textbf{Drill}} \\ \cline{2-19} 
                                                                                       & \textbf{Text}   & \textbf{Parquet}   & \textbf{Text}     & \textbf{Parquet}    & \textbf{Text}   & \textbf{Parquet}  & \textbf{Text}   & \textbf{Parquet}   & \textbf{Text}     & \textbf{Parquet}    & \textbf{Text}   & \textbf{Parquet}  & \textbf{Text}   & \textbf{Parquet}   & \textbf{Text}     & \textbf{Parquet}    & \textbf{Text}   & \textbf{Parquet}  \\ \hline
\textbf{Q1}                                                                            & 37.68           & 37.79              & \textbf{24.8}              & \textbf{4.86}                & 33.33           & 18.06             & 71.09           & 73                 & \textbf{46.23}             & \textbf{6.56}                & 63.81           & 26.9              & 142.43          & 138.6              & \textbf{78.46}            & \textbf{22.56}               & 132.33          & 50.7              \\ \hline
\textbf{Q2}                                                                            & \textbf{3.59}            & \textbf{2.58}               & 43.4              & 16.03               & Failed          & Failed            & \textbf{5.56}            & \textbf{3.43}               & 72.93             & 23.86               & Failed          & Failed            & \textbf{8.73}            & \textbf{4.1}                & 152.5             & 33.2                & Failed          & Failed            \\ \hline
\textbf{Q3}                                                                            & \textbf{9.08}            & 7.34               & 28.7              & \textbf{7.3}                 & 52.16           & 16.5              & \textbf{15.66}           & 12.66              & 54.5              & \textbf{11.5}                & 92.46           & 23.7              & \textbf{28.33}           & 24.2               & 112.93            & \textbf{22.13}               & 189.46          & 40.46             \\ \hline
\textbf{Q4}                                                                            & \textbf{7.94 }           & \textbf{7.31}               & 44.2              & 25.56               & 54.76           & 18.36             & \textbf{13.76}           & \textbf{12.66}              & 79.8              & 46.46               & 94.63           & 29.1              & \textbf{28.5}            & \textbf{16.66}              & 171.4             & 94                  & 197             & 44.9              \\ \hline
\textbf{Q5}                                                                            & \textbf{13.3}            & \textbf{10.44}              & 46.1              & 25.16               & 50.23           & 13.8              & \textbf{23.9}            & 26.63              & 97.13             & 35.1                & 95.83           & \textbf{18.33}             & \textbf{44.63}           & 52.2               & 151.06            & 62                  & 198.73          & \textbf{37.16}             \\ \hline
\textbf{Q6}                                                                            & \textbf{3}               & \textbf{1.81}               & 21.63             & 2.26                & 29.93           & 5.26              & \textbf{5.86}            & \textbf{1.84}               & 36.76             & 2.7                 & 57.66           & 6.73              & \textbf{11.16}           & \textbf{2.8}                & 74.03             & 3.9                 & 125.73          & 12.86             \\ \hline
\textbf{Q7}                                                                            & \textbf{13.47}           & \textbf{13.29}              & 75.81             & 22.4                & 52.1            & 17.96             & \textbf{22.2}            & \textbf{21.76}              & 144.33            & 61.56               & 100.4           & 32.83             & \textbf{42.16}           & \textbf{44.6}               & 244.43            & 65.06               & 199.5           & 66.73             \\ \hline
\textbf{Q8}                                                                            & \textbf{5.68}            & \textbf{3.49}               & 47.33             & 18.86               & 42.23           & 5.53              & \textbf{10.6}            & \textbf{6.7}                & 67                & 41.5                & 87.2            & 11.46             & \textbf{21.73}           & \textbf{11.53}              & 128.4             & 78.73               & 208.56          & 22.93             \\ \hline
\textbf{Q9}                                                                            & \textbf{14.79}           & \textbf{12.29}              & 54.26             & 30.36               & 46.7            & 12.4              & \textbf{23.53}           & \textbf{17.73}              & 95.7              & 66.73               & 112.3           & 28.1              & \textbf{40.13}           & \textbf{31.06}              & 181.13            & 125                 & 231.33          & 61.7              \\ \hline
\textbf{Q10}                                                                           & \textbf{7.27}            & \textbf{5.68}               & 32.83             & 8.56                & 40.73           & 12.86             & \textbf{11.6}            & \textbf{9.13}               & 62.26             & 10.76               & 79              & 19.06             & \textbf{20.73}           & \textbf{15.5}               & 110.13            & 32.56               & 168.86          & 40.46             \\ \hline
\textbf{Q11}                                                                           & Failed          & Failed             & 33.23             & 21.33               & \textbf{11.93}           & \textbf{1.66}              & Failed          & Failed             & 51.46             & 28.76               & \textbf{19.76}           & \textbf{1.8}               & Failed          & Failed             & 77.96             & 37.63               & \textbf{32.66}           & \textbf{3.66}              \\ \hline
\textbf{Q12}                                                                           & \textbf{6.58}            & 4.97               & 26.76             & \textbf{4.76}               & 37.43           & 18.1              & \textbf{11.23}           & 7.96               & 48.1              & \textbf{6.5}                 & 76.7            & 28.3              & \textbf{21.03}           & 12.93              & 93.56             & \textbf{10.43}               & 157.6           & 58.3              \\ \hline
\textbf{Q13}                                                                           & \textbf{9.01}            & 10.75              & 16.43             & \textbf{9.26}                & Failed          & 9.36              & \textbf{17.86}           & 20.46              & 22.96             & 15.23               & Failed          & \textbf{9.4}               & 34.93           & 36                 & \textbf{34.86}             & 25.73               & Failed          & \textbf{22.9}             \\ \hline
\textbf{Q14}                                                                                    & \textbf{4.44}            & 3.88               & 23.26             & \textbf{3.73}                & 30.2            & 4.5               & \textbf{7.5}             & \textbf{4.4}                & 39.9              & 4.96                & 61.53           & 8.83              & \textbf{12.93}           & 8.3                & 72.9              & \textbf{7.23}                & 134.76          & 18.33             \\ \hline
\textbf{Q15}                                                                           & \textbf{11.26}           & 8.4                & 49.5              & \textbf{4.96}                & 30.6            & 6                 & \textbf{18.4}           & 9.92               & 77.96             & \textbf{7.36}                & 61.99           & 13.26             & \textbf{31.88}           & 16.21              & 147.06            & \textbf{12.03}               & 129.06          & 22.26             \\ \hline
\textbf{Q16}                                                                           & \textbf{6.37}            & \textbf{6.32}               & 68.93             & 81                  & Failed          & 8.46              & \textbf{6.16}            & \textbf{8.36}               & 136.76            & 152                 & Failed          & 17.06             & \textbf{12.75}           & \textbf{12.96}              & 286.93            & 371.96              & Failed          & 25.3              \\ \hline
\textbf{Q17}                                                                           & \textbf{29.65}           & 31.19              & 53.8              & 17.06               & 95.26           & \textbf{14.96}             & \textbf{61.23}           & 64.73              & 99.93             & 41.93               & 182.26          & \textbf{24.13}             & \textbf{125.76}          & 96.74              & 202.06            & 87.06               & 361.63          & \textbf{43.76}             \\ \hline
\textbf{Q18}                                                                           & \textbf{17.8}            & 17.82              & 63.6              & \textbf{15.13}               & 98.76           & 25.93             & \textbf{36.24}           & 30.2               & 110.83            & \textbf{27.23}               & 226.36          & 48.16             & \textbf{72.93}           & \textbf{47.03}              & 201.9             & 51.5                & 478.06          & 124.83            \\ \hline
\textbf{Q19}                                                                           & 49.69           & 52.22              & \textbf{22.96}             & \textbf{4.46}                & Failed          &        Failed           & 105.73          & 114.53             & \textbf{40.63}             & \textbf{5.36}                & Failed          & Failed            & 211.13          & 209.13             & \textbf{74.46}             & \textbf{8.8}                 & Failed          & Failed            \\ \hline
\textbf{Q20}                                                                           & \textbf{8.54}            & \textbf{4.53}               & 38.26             & 21.43               & 45.13           & 9.86              & \textbf{15.24}           & \textbf{7.46}               & 57.26             & 19.1                & 91.03           & 18.4              & \textbf{29.1}            & \textbf{13.16}              & 110.86            & 26.73               & 171.96          & 27.63             \\ \hline
\textbf{Q21}                                                                           & \textbf{24.19}           & \textbf{23.15}              & 131.7             & 83.63               & Failed          & Failed            & \textbf{46.8}            & \textbf{44.83}              & 265.66            & 172.8               & Failed          & Failed            & \textbf{95.83}           & \textbf{89}                 & 583.5             & 364.43              & Failed          & Failed            \\ \hline
\textbf{Q22}                                                                           & \textbf{3.36}            & \textbf{2.86}               & 22.86             & 13.3                & Failed          & Failed            & \textbf{5.16}            & \textbf{5.13}               & 30.7              & 17.26               & Failed          & Failed            & \textbf{8.7}             & \textbf{8.33}               & 48.23             & 35.66               & Failed          & Failed            \\ \hline
\textbf{AM}                                                                            & --              & --                 & 44.11             & 20.06                & --              & --                & --              & --                 & 79.04             & 36.6                & --              & --                & --              & --                 & 151.76            & 71.74               & --              & --                \\ \hline
\textbf{\begin{tabular}[c]{@{}c@{}}AM--Q\{2,11,13,\\ 16,19,21,22\}\end{tabular}}       & 12.86           & 11.73              & 41.37             & 14.15               & 50.15           & 12.76             & 23.64           & 21.29              & 72.84             & 26.58               & 100.46          & 21.23             & 45.82           & 36.97              & 134.77            & 47.75               & 209.07          & 42.51             \\ \hline
\textbf{\begin{tabular}[c]{@{}c@{}}Normalized AM--Q\\\{2,11,13,16,19,21,22\}\end{tabular}}       & 1.26          & 1              & 5.83             & 1.87               & 6.75           & 1.68             & 2.25          &   1.65            & 9.95             & 3.15               & 13.4          & 2.79             & 4.27           & 2.86              & 18.86            & 5.65               & 28.41          & 5.8             \\ \hline
\textbf{\begin{tabular}[c]{@{}c@{}}Text (AM) over\\ Parquet (AM)\end{tabular}}       & \multicolumn{2}{c|}{1.26}  & \multicolumn{2}{c|}{3.11}    & \multicolumn{2}{c|}{4.01}            & \multicolumn{2}{c|}{1.36}              & \multicolumn{2}{c|}{3.15}            & \multicolumn{2}{c|}{4.8}           & \multicolumn{2}{c|}{1.49}             & \multicolumn{2}{c|}{3.33}            & \multicolumn{2}{c|}{4.89}             \\ \hline
\end{tabular}
}
\end{table*}
\section{TPC-H Benchmark Experiments}
\label{sec:tpch}
Next, we utilize the TPC-H benchmark to evaluate and compare the performance of Impala, Drill and Spark SQL systems. We use TPC-H since its workload comprises of a wide range of complex analytical tasks that thoroughly test query expressiveness, optimizer quality and query execution engine efficiency in the examined systems. We evaluate each system with the data stored in, both the text and the parquet storage formats. The parquet format enables analytical systems to achieve improved disk I/O efficiency by allowing reads to skip unnecessary columns, and improved storage efficiency through compression and encoding schemes. To evaluate the \textit{size-up} characteristic of each examined system, we perform experiments for three scale-factors (SFs) : 125, 250, and 500. Note that SF denotes the database size (in GB) in the text format. We exclude P-HBase from the TPC-H experiments since more than 90\% of the benchmark queries require evaluation of one or more joins to compute the output RS. However, the P-HBase execution engine architecture with client coordinated shuffle is not apt for join heavy workloads and results in orders of magnitude slower query performance as compared to the other evaluated systems. We use the same experiment setup for each evaluated system as described in Section \ref{sec:ex-setup}. 
\begin{table}[h]
\centering
\caption{Data preparation times (seconds) in systems for the TPC-H benchmark}
\label{load-tpch}
\scalebox{.56}{
\begin{tabular}{|c|c|c|c|c|c|c|c|c|c|c|}
\hline
\multirow{3}{*}{\textbf{\begin{tabular}[c]{@{}c@{}}TPC-H \\ Scale\\ Factor\end{tabular}}} & \multirow{3}{*}{\textbf{\begin{tabular}[c]{@{}c@{}}Insert \\ HDFS\end{tabular}}} & \multicolumn{6}{c|}{\textbf{Impala}}                                                                                                                                                                                                                                                                        & \multicolumn{3}{c|}{\textbf{Spark SQL and Drill}}      \\ \cline{3-11} 
                                                                                          &                                                                                  & \multicolumn{3}{c|}{\textbf{Text}}                                                                                                                   & \multicolumn{3}{c|}{\textbf{Parquet}}                                                                                                                & \textbf{Text}  & \multicolumn{2}{c|}{\textbf{Parquet}} \\ \cline{3-11} 
                                                                                          &                                                                                  & \textbf{\begin{tabular}[c]{@{}c@{}}Load \\ Tables\end{tabular}} & \textbf{\begin{tabular}[c]{@{}c@{}}Compute \\ Stats\end{tabular}} & \textbf{Total} & \textbf{\begin{tabular}[c]{@{}c@{}}Load \\ Tables\end{tabular}} & \textbf{\begin{tabular}[c]{@{}c@{}}Compute \\ Stats\end{tabular}} & \textbf{Total} & \textbf{Total} & \textbf{Convert}   & \textbf{Total}   \\ \hline
\textbf{125}                                                                              & 957                                                                              & 31.4                                                            & 118.4                                                             & 1106.8         & 120.2                                                           & 127.7                                                             & 1236.3         & 957            & 134.8              & 1091.8           \\ \hline
\textbf{250}                                                                              & 1900                                                                             & 32.8                                                            & 214.6                                                             & 2147.4         & 205.9                                                           & 247.2                                                             & 2385.9         & 1900           & 212.3              & 2112.3           \\ \hline
\textbf{500}                                                                              & 3898                                                                             & 33.4                                                            & 415                                                               & 4346.4         & 354.1                                                           & 441.9                                                             & 4727.4         & 3898           & 328                & 4226             \\ \hline
\end{tabular}
}
\end{table}
\vspace*{-6mm}
\subsection{Hardware and Software Configuration} 
Our experiment testbed comprises of 20 \textit{worker} VMs, 1 \textit{client} VM and 1 \textit{master} VM (see Section \ref{sec:hw-conf} for VM instance descriptions). We use the same software configuration for each evaluated system as described in Section \ref{sec:sw-conf}.
\subsection{Data Preparation}
For the text format, we use the same data preparation steps in each evaluated system as described in Section \ref{wdab-prep}. For the parquet format, we take different steps in the Impala and the Spark SQL systems. In Spark SQL, for each TPC-H table, we use a script to first read the text files stored in HDFS into a \textit{rdd}, then convert the \textit{rdd} into a \textit{data frame} and finally save the \textit{data frame} back into HDFS in the parquet format. Created parquet files are then queried in, both Drill and Spark SQL systems. In Impala, we first create the schema for parquet tables and then load parquet tables using the text tables. Next, we utilize the \textsc{compute stats} command to collect statistics for each parquet table. We use \textit{Snappy} compression with parquet format in each evaluated system. Table \ref{load-tpch} shows the data preparation times for each evaluated system at TPC-H scale factors 125, 250 and 500. The DB size in the parquet format at scale factors 125, 250 and 500 is 39.9GB, 79.8GB and 168.1GB respectively. The data preparation in each system for both the text and the parquet formats increases proportionately with the increase in the data size, exhibiting good \textit{size-up} property.

\begin{figure*}[t]
\hspace*{.15 in}
\includegraphics[clip=true, trim=149 8 80 37,scale=.37]{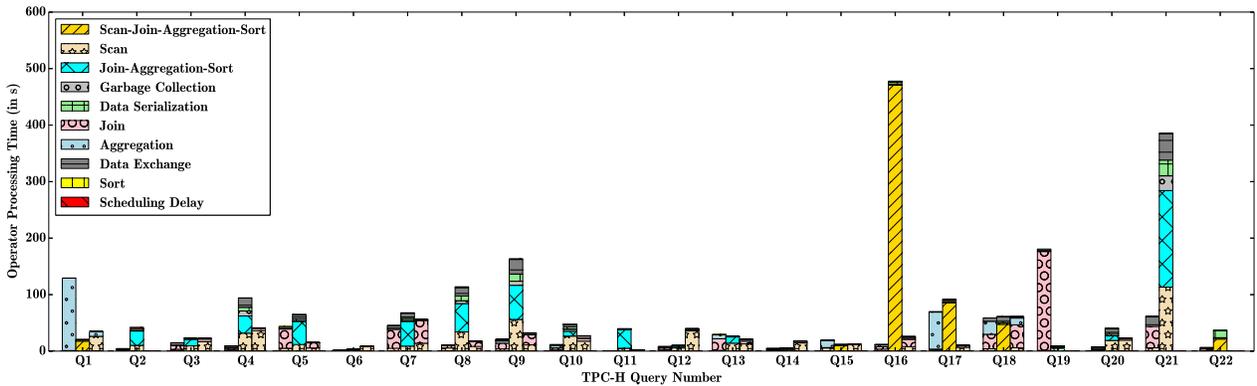} 
\centering
\caption{\small The breakdown of query RT into aggregated processing time for each operator type in evaluated systems. The TPC-H scale factor is 500 and the storage format is parquet. For each query, the left bar, the middle bar and the right bar represent Impala, Spark SQL and Drill systems, respectively.}
\label{fig:parq-aggr}
\end{figure*}
\subsection{Experiment Results}
Table \ref{tpch-numbers} presents the RT of TPC-H queries in each evaluated system for the text and the parquet storage formats at scale factors 125, 250, and 500. Again, the standard error of the mean query RT is minimal in evaluated systems; hence, we exclude it from the presentation of the results. Table \ref{tpch-numbers} also shows the arithmetic mean (AM) of the RT of all benchmark queries for each storage format, SF, and evaluated system combination.

Only 15 TPC-H queries could be evaluated in each system. AM--Q\{2,11,13,16,19,21,22\} represents the AM of the RT of all benchmark queries except Q2, Q11, Q13, Q16, Q19, Q21, Q22. The query optimizer in Impala failed to plan for Q11. Drill exhibits minimal query expressiveness with six failed queries in the two storage formats. Queries Q2, Q19, Q21 and Q22 failed for both storage formats in Drill with a server side ``Drill Remote Exception'', whereas queries Q13 and Q16 failed with the same error for the text storage format only.

We use the normalized AM--Q\{2,11,13,16,19,21,22\} (see Table \ref{tpch-numbers}) to carry out an overall performance comparison of evaluated systems for the text and the parquet storage formats at scale factors 125, 250, and 500. In the text format, Impala is the fastest (4.41x -- 6.65x) and Drill is the slowest (1.15x -- 6.65x), across all evaluated scale factors. In the parquet format, although Drill is nearly 1.1x faster than Spark SQL at smaller scale factors (SF 125 and SF 250), Spark SQL marginally outperforms (1.02x) Drill for the largest evaluated scale factor (SF 500). Impala is the fastest (1.68x -- 2.0x) system in the parquet format, across all evaluated scale factors. In contrast with the text format, parquet format results exhibit interesting query performance trade-offs with each system outperforming the other two systems for a subset of TPC-H queries.

In the subsequent sections, we analyze the query execution profiles to gain an insight into the optimizer characteristics and the execution engine efficiency in each evaluated system.
\begin{figure*}[ht]\centering
\hspace*{-2.0 in}
\begin{minipage}[t]{0.24\textwidth}\centering
\subfigure{
\includegraphics[clip=true, trim=20 70 30 14, scale=.3]{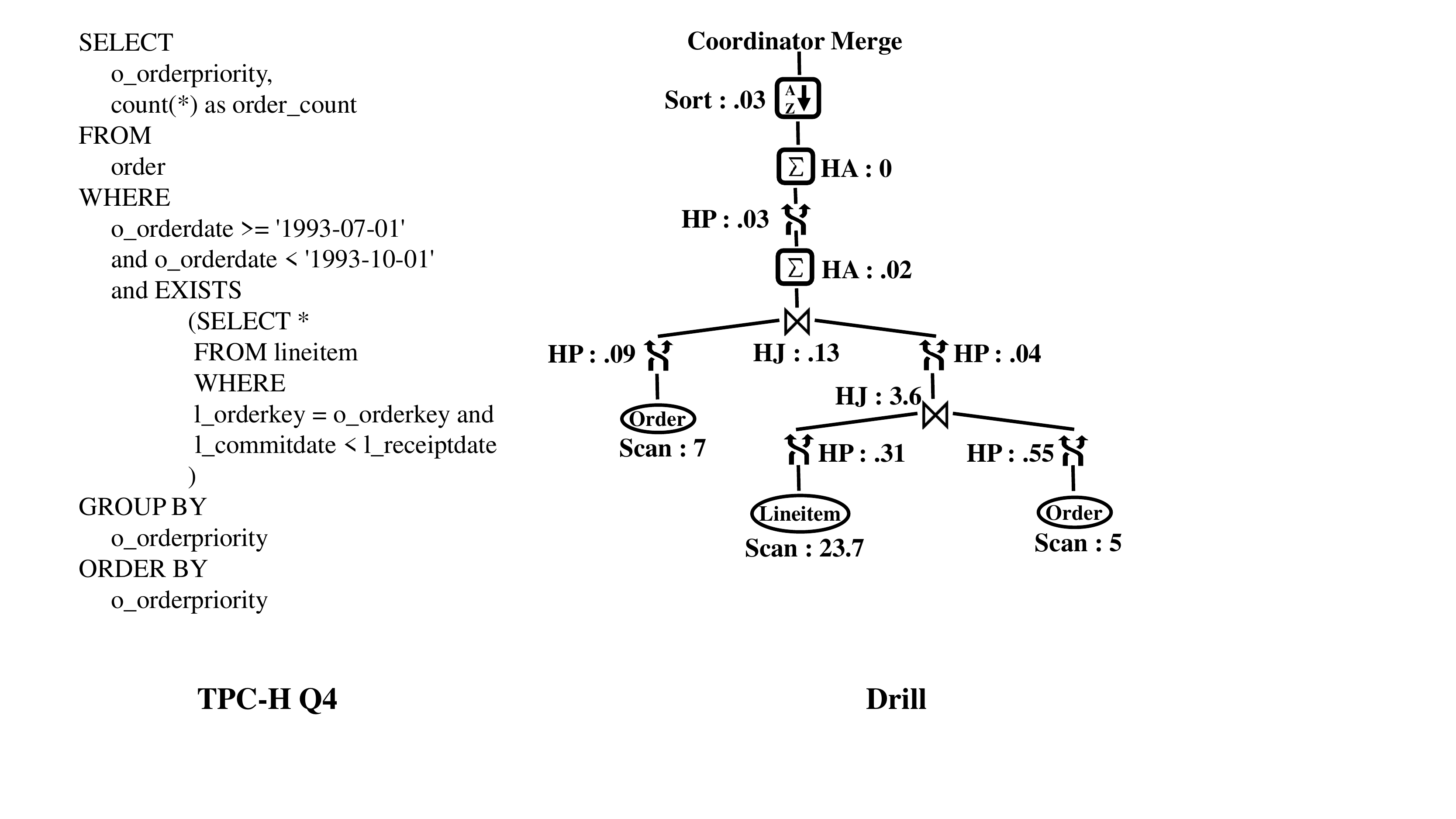} 
\label{fig:q4-1}
}
\end{minipage}
\hspace*{1.8 in}
\begin{minipage}[t]{0.24\textwidth}\centering
\subfigure{
\includegraphics[clip=true, trim=30 70 30 14, scale=0.3]{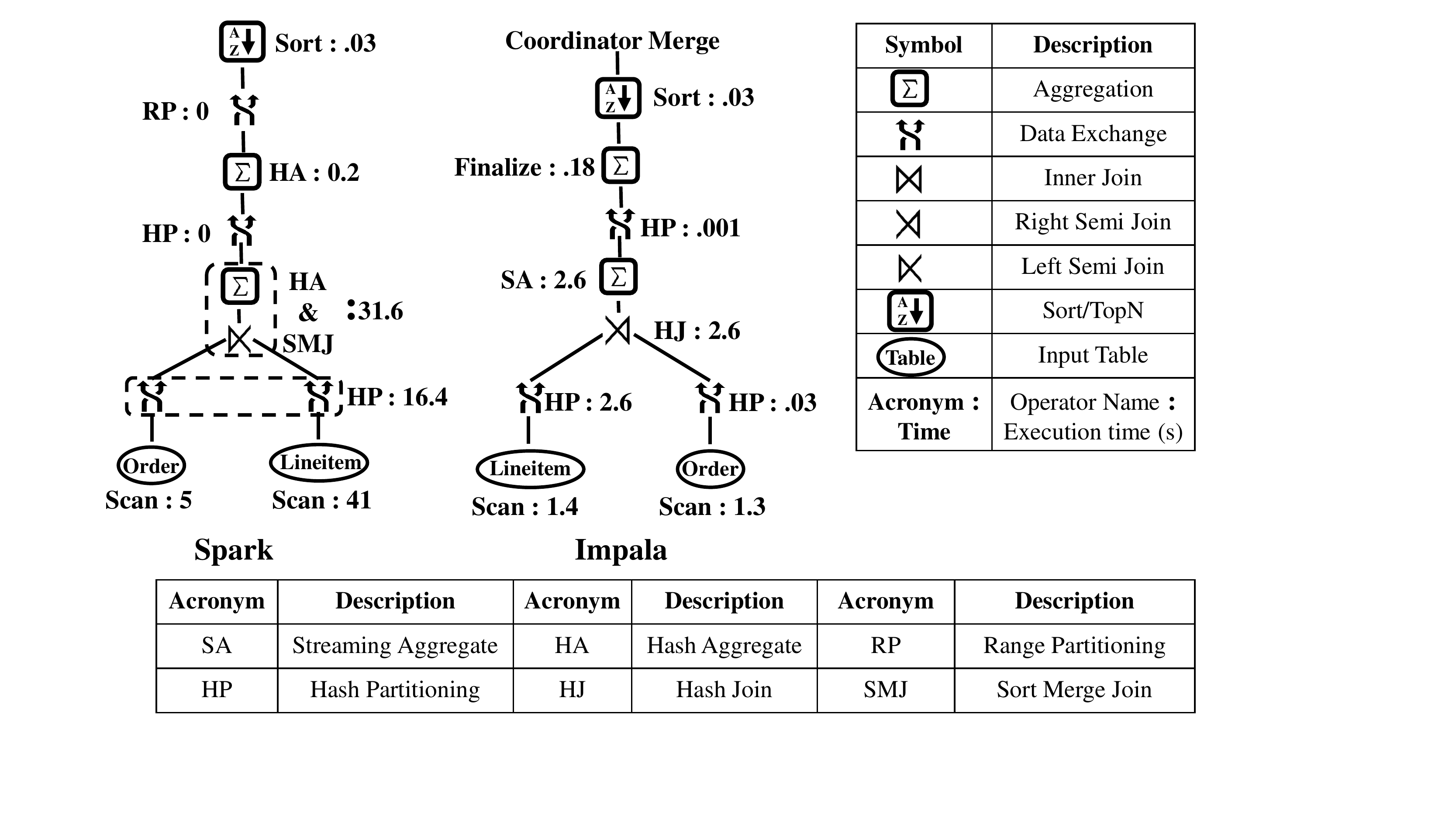} 
\label{fig:q4-2}
}
\end{minipage}
\caption{\small Query text and execution plans with profiled operator times in evaluated system for TPC-H query 4. The storage format is parquet and the SF is 500}
\label{fig:q4}
\end{figure*}
\subsubsection{Execution Time Breakdown}
In this section we present the breakdown of query RT into aggregated execution time for each operator type to understand execution characteristics in evaluated systems. We perform this analysis for the largest evaluated scale factor (SF 500) in the parquet format (see Section \ref{parq-vs-txt} for parquet vs. text comparison). Figure \ref{fig:parq-aggr} depicts the execution time breakdown for the TPC-H benchmark queries in evaluated systems.

\textbf{\textsc{impala.}} Impala primarily utilizes HJ and SA operators to perform join and aggregation operations, respectively. The query RT is dominated by scan, join and aggregation operator times in Impala. On an average, the join and aggregation operator times are 35\% and 25\% of the query RT, respectively. Use of a single CPU core to perform the join and the aggregation operations (identified in previous work \cite{floratou2014} as well) combined with the choice of SA operator to perform the grouping aggregation, results in sub-optimal CPU and memory resource usage and is the primary performance bottleneck in Impala. Although, on an average, the scan operator time is 18\% of the query RT, Impala exhibits the most efficient disk I/O sub-system among all evaluated systems. The average data-exchange operator time is 6\% of the query RT, demonstrating efficient network I/O subsystem. Query 19 RT is relatively high in Impala since predicates are evaluated during the join operation instead of being pushed down to the scan operation.

\textbf{\textsc{drill.}} Drill mainly uses HA and SA operators to perform grouping and non-grouping aggregation operations, respectively. In addition, HJ represents the primary join operator in Drill. The scan operator contributes the maximum (on an average 42\%) to the query RT in Drill. The total scan operator time in Drill is nearly 4.5x as compared to Impala for all benchmark queries that completed in both systems. Although, on an average, the join operator time in Drill is 21\% of the query RT, the HJ operator choice combined with an efficient operator implementation results in lowest total join operator time for all benchmark queries among all evaluated systems. Drill exhibits high scheduling overhead with average time being 13\% of the query RT; however, the data-exchange operator shows notable efficiency with average time being 4\% of the query RT.

\textbf{\textsc{spark sql.}} Recall that multiple query plan operators may be mapped to a single DAG stage in Spark SQL (see Section \ref{sec:profiling}). Hence, for queries that perform: 1) partial-aggregation and join, and/or 2) final-aggregation and sort operations in a single stage, we present the sum of join, aggregation, and sort operator times, denoted as JAS. Also, for queries that perform scan and partial-aggregation operations in a single stage, we present the sum of scan and JAS operation times.

Spark SQL largely utilizes SMJ and HA operators to perform join and aggregation operations, respectively. On an average, the scan and the JAS operations contribute 42\% and 46\% to the query RT, respectively (based on the 15 TPC-H queries that perform scan and JAS operations in separate stages). The joins are expensive in Spark SQL due to use of SMJ operator that performs a costly sort operation on both join inputs before combining the matching records. On an average, the GC time is 7\% of the query RT and scan operation represents the principal source of GC overhead. Although the average data-exchange operator time is 7\% of the query RT, the network data transfer performance in Spark SQL is at least 3x slower as compared to other evaluated systems based on the total benchmark data-exchange time.
\subsubsection{Correlated Sub-query Execution in Drill}
\label{correlated}
In this section we discuss correlated sub-query execution characteristic in Drill through an example TPC-H query (Q4). In the case of correlated sub-queries with one or more filters on the outer table, Drill optimizer generates a query execution plan that first performs a join of the outer and inner table to filter the inner table rows for which the join key does not match with the join key of the filtered outer table. The filtered inner table rows are then joined with the outer table rows to generate the output RS. We also compare the Drill query execution with the query execution in Impala and Spark SQL systems. Figure \ref{fig:q4} depicts the query text, execution plans and the profiled operator times in evaluated systems for the TPC-H query 4. Note that the storage format is parquet and scale factor is 500.

\textbf{\textsc{drill.}} The Order table is scanned (o\_orderkey, o\_orderdate), filtered (o\_orderdate \textgreater= `1993-07-01' and o\_orderdate \textless `1993-10-01') and hash-partitioned on the o\_orderkey. Similarly, Lineitem table is scanned (l\_orderkey, l\_commitdate, l\_receiptdate), filtered (l\_commitdate \textless l\_receiptdate) and hash-partitioned on the l\_orderkey. Next, tuples from the Order and the Lineitem partitions are inner joined using the HJ operator and the intermediate RS is hash-partitioned on the o\_orderkey. This join operation reduces the number of Lineitem rows that are shuffled across the cluster nodes. Next, Order table is scanned (o\_orderkey, o\_orderdate, o\_orderpriority), filtered (o\_orderdate \textgreater= `1993-07-01' and o\_orderdate \textless `1993-10-01') and hash-partitioned on the o\_orderkey for the second time. Then, tuples from the intermediate RS and Order partitions are inner joined using the HJ operator. Subsequently, the results are partially hash-aggregated and hash-partitioned on the grouping attribute (o\_orderpriority) to enable final hash-aggregation. Finally, sorted results are merged in the coordinator node. 

\begin{figure*}[th]
\hspace*{.15 in}
\includegraphics[clip=true, trim=149 8 80 37,scale=.37]{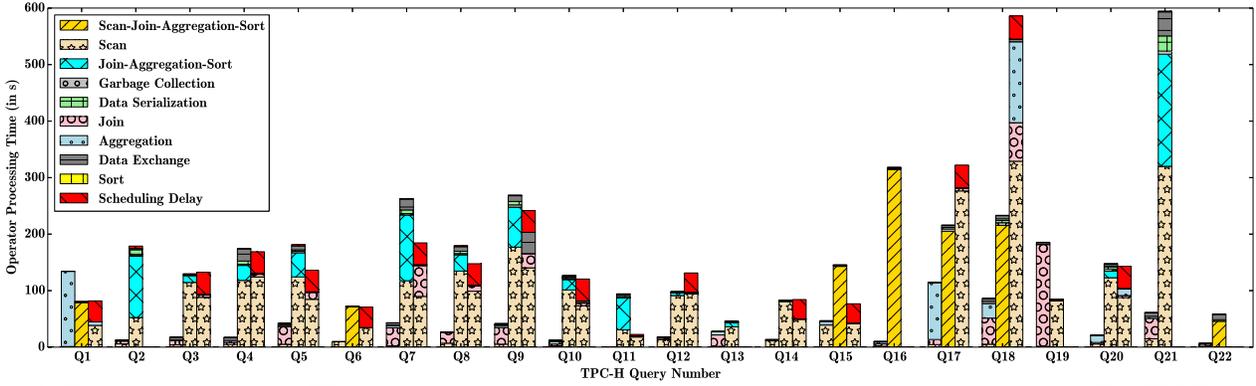}
\centering
\caption{\small The breakdown of query RT into aggregated processing time for each operator type in evaluated systems. The TPC-H scale factor is 500 and the storage format is text. For each query, the left bar, the middle bar and the right bar represent Impala, Spark SQL and Drill systems, respectively.}
\label{fig:text-aggr}
\end{figure*}
The first join between the Lineitem and the Order table reduces the data that are partitioned across the cluster nodes. However, as shown in Figure \ref{fig:q4}, scan operation is the primary performance bottleneck in Drill. In addition, using the same plan with text data worsens the performance since all columns in the Order table are scanned twice during query execution.

\textbf{\textsc{spark sql.}} The Lineitem and the Order tables are scanned and hash-partitioned on the join keys (o\_orderkey, l\_orderkey). The tuples from the Order and Lineitem table partitions are then left-semi joined using the SMJ operator. The results are then partially hash-aggregated and hash-partitioned on the grouping attribute (o\_orderpriority) to enable final hash-aggregation. The aggregated results are then range partitioned and sorted to generate the output RS. Although only three columns need to be scanned from both tables in the parquet format, due to a bug in the plan generation for queries with \textit{exists} clause, all columns are scanned in both tables. As a result, scan operation is very costly for both tables. In addition, since relevant columns are projected after the join operation, the data-exchange operation and the sort operation in SMJ (required disk spill) are expensive as well.  

\textbf{\textsc{impala.}} Similar to Spark SQL, Impala scans and hash-partitions the Lineitem and the Order tables on the join keys (o\_orderkey, l\_orderkey). The tuples from the Lineitem and the Order table partitions are then right-semi joined using the HJ operator. The join results are then partially hash-aggregated and hash-partitioned on the grouping attribute (o\_orderpriority) to enable final hash-aggregation. The aggregated results are then sorted and merged in the coordinator to produce the output RS. A simple and effective query execution plan combined with an efficient disk I/O subsystem enables Impala to outperform other systems by at least a factor of 2.
\subsubsection{Parquet versus Text Performance}
\label{parq-vs-txt}
In this section we evaluate the query performance differences between the text and parquet storage formats in each evaluated system. The last row in Table \ref{tpch-numbers} shows the ratio of overall text to parquet performance in each system for TPC-H scale factors 125, 250 and 500. These numbers were computed using the normalized AM--Q\{2,11,13,16,19,21,22\} values. Drill and Impala exhibit the maximum (4.01x -- 4.89x) and minimum (1.26x -- 1.49x) performance speed-up, respectively from the text to the parquet format. Also, as we increase the data size, the speed-up factor between the two formats increases in each system. Next, we use results from the largest TPC-H scale factor (SF 500) to understand the reasons for the performance difference between the two storage formats in evaluated systems. For each system, figures \ref{fig:parq-aggr} and \ref{fig:text-aggr} show the per operator total time spent by a query in the parquet and text storage formats, respectively.
  
\textbf{\textsc{impala.}} Impala explicitly collects statistics on the data and generates the same query execution plan in both formats. On an average, the query scan operator time in the text format is 3.2x the parquet format. The input and output data sizes for a join operator remain same in both storage formats; however, on an average, the HJ operator time in the text format is 3x as compared to the parquet format. Excluding query 17 for which the partial-aggregation operator time increases by 60\% from the parquet to the text format, the total aggregation operator time for all benchmark queries shows nominal difference (less than 5\%) between the two formats. The sort and the data-exchange operator times exhibit minimal variance between two storage formats. Hence, scan and HJ operators are the primary contributors to the increase in query RT in the text format as compared to the parquet format.

\textbf{\textsc{drill.}} Drill generates the same query execution plan for both storage formats. The join, aggregation, sort and data-exchange operators exhibit nominal difference in the processing time between the two storage formats. The scan operator is the principal performance bottleneck in the text format in Drill, since on an average, the scan operator time in the text format is 12x as compared to the parquet format.
\begin{table}[h]
\centering
\vspace*{1mm}
\caption{Size-up property evaluation in each system. SF denotes scale factor.}
\label{sizeup}
\scalebox{.52}{
\begin{tabular}{|c|c|c|l|c|c|c|c|}
\hline
\multirow{2}{*}{\textbf{Storage Format}} & \multicolumn{3}{c|}{\textbf{Impala}}                                            & \multicolumn{2}{c|}{\textbf{Spark SQL}}                   & \multicolumn{2}{c|}{\textbf{Drill}}                       \\ \cline{2-8} 
                                         & \textbf{SF250 / SF125}  & \multicolumn{2}{c|}{\textbf{SF500 / SF250}} & \textbf{SF250 / SF125} & \textbf{SF500 / SF250} & \textbf{SF250 / SF125} & \textbf{SF500 / SF250} \\ \hline
\textbf{Text}                            & 1.78                         & \multicolumn{2}{c|}{1.89}                        & 1.76                        & 1.89                        & 1.99                           & 2.1                        \\ \hline
\textbf{Parquet}                         & 1.65                         & \multicolumn{2}{c|}{1.73}                        & 1.68                        & 1.79                        & 1.66                        & 2.07                         \\  \hline
\end{tabular}
}
\end{table}
\vspace*{-3mm}

\textbf{\textsc{spark sql.}} Spark SQL generates query execution plans with the same join order in both formats. To query the text data, the Spark SQL optimizer harnesses SMJ operator to perform all the joins in the execution plan. However, for the parquet data, small tables (region and nation) are exchanged using the broadcast mechanism and the HJ operator is utilized to join tuples from the small table and the other join input. The joins performed using the same operator (SMJ) show nominal difference between the two storage formats. On an average, the scan operator time in the text format is 8.7x as compared to the parquet format. Note, remaining operations (data-exchange, GC, etc.) show insignificant difference in processing time between the two storage formats.
\begin{table*}[th]
\centering
\caption{Summary and classification of related works.}
\label{related-works}
\scalebox{0.7}{
\hspace*{.45 in}
\begin{tabular}{|c|c|c|c|c|c|lll}
\cline{1-6}
\textbf{Works}                                 & \textbf{Query Engines Compared}                                                                                               & \textbf{Benchmarks}                                                              & \textbf{Cluster Specification}                                              & \textbf{DB Size (Max.)} & \textbf{Experiment Metrics}                                    &  &  &  \\ \cline{1-6}
\textbf{Pavlo et al. \cite{pavlo}}           & \begin{tabular}[c]{@{}c@{}}MR, DBMS-X, Vertica\end{tabular}    & \begin{tabular}[c]{@{}c@{}}WDA \end{tabular}  & \begin{tabular}[c]{@{}c@{}}100 nodes (max), Private Cluster\end{tabular} & 2.1 TB                  & \begin{tabular}[c]{@{}c@{}}RT, Scale Up, Speed Up\end{tabular} &  &  &  \\ \cline{1-6}
\textbf{Floratou et al. \cite{floratou2012}} & Hive-MR, SQL Sever PDW                                                                                                        & TPC-H                                                                           & 16 nodes, Private Cluster                                                   & 16 TB                   & RT, Size Up                                                       &  &  &  \\ \cline{1-6}
\textbf{Floratou et al. \cite{floratou2014}} & Hive-Tez, Hive-MR, Impala                                                                                                     & TPC-H, TPC-DS derived                                                           & 21 nodes, Private Cluster                                                   & 1 TB                    & RT                                                                &  &  &  \\ \cline{1-6}
\textbf{\cite{amp}}                          & Redshift, Shark, Impala, Hive-MR, Hive-TEZ                                                                                    & \begin{tabular}[c]{@{}c@{}}AMP Lab BigData Benchmark (ALBB)\end{tabular}     & 5 nodes, Amazon EC2                                                         & 127.5 GB                & RT                                                                &  &  &  \\ \cline{1-6}
\textbf{Wouw et al. \cite{icpe}}             & Hive-MR, Shark, Impala                                                                                                        & ALBB, Real World                                                                & 5 nodes, Amazon EC2                                                         & 523 GB                  & RT, Speed Up, Size Up                                             &  &  &  \\ \cline{1-6}
\textbf{Pirzadeh et al. \cite{bigfun}}       & \begin{tabular}[c]{@{}c@{}}AsterixDB, System-X, Hive-MR, MongoDB\end{tabular}                                              & \begin{tabular}[c]{@{}c@{}}BigFun Micro Benchmark\end{tabular}               & 10 nodes, Private Cluster                                                   & 800 GB                  & RT, Scale Up                                                      &  &  &  \\ \cline{1-6}
\textbf{Shi et al. \cite{titans}}            & Spark, MR                                                                                                                     & \begin{tabular}[c]{@{}c@{}}Word Count, Page Rank, K-Means, Sort\end{tabular} & 4 nodes, Private Cluster                                                    & 500 GB                  & RT                                                                &  &  &  \\ \cline{1-6}                              
\textbf{Our Work}                              & \begin{tabular}[c]{@{}c@{}}\textbf{P-HBase, Drill, Spark SQL, Impala}\end{tabular}                                            & \textbf{WDA, TPC-H}                                                                     & \begin{tabular}[c]{@{}c@{}}\textbf{21 nodes (max), Amazon EC2}\end{tabular}       & \textbf{500GB}                & \begin{tabular}[c]{@{}c@{}}\textbf{RT, Scale Up, Size Up}\end{tabular}  &  &  &  \\ \cline{1-6}
\end{tabular}
}\end{table*}
\subsubsection{Size-up Characteristic Evaluation}
In this section we assess the \textit{size-up} behavior in the evaluated systems as we increase the TPC-H scale factor in multiples of 2 between 125 and 500. Table \ref{sizeup} presents the ratio of overall performance at consecutive scale factors in evaluated systems for both storage formats. These numbers were computed using the normalized AM--Q\{2,11,13,16,19,21,22\} values. 

\textbf{\textsc{impala.}} Impala exhibits sub-linear size-up behavior for both storage formats. On an average, the join, aggregation and data-exchange operator times double as the database size is doubled. However, the scan operator time exhibits sub-linear increase, resulting in the sub-linear size-up behavior in Impala.  

\textbf{\textsc{drill.}} With the increase in the database size, the optimizer's join procedure selection in Drill favors hash-partitioned HJ as compared to the broadcast HJ. Hence, in the parquet format, Drill chooses more broadcast HJs for scale factor 125 (DB size -- 39.9 GB) as compared to the scale factor 250 (DB size -- 79.8 GB). Since broadcast HJs exhibit higher execution times in comparison with the hash-partitioned HJs in Drill, sub-linear size-up behavior is observed as the scale factor is doubled from 125 to 250 in the parquet format.

\textbf{\textsc{spark sql.}} Spark SQL shows sub-linear size-up behavior for both storage formats. In the text format, although the JAS operation time reduces marginally as the database size is doubled, the reduction in scan operation time is primarily responsible for the sub-linear size-up behavior. In the parquet format, the reduction in both scan and JAS operation times is accountable for the sub-linear size-up behavior.  
\section{Related Work}
Table \ref{related-works} presents a classification of the related works. In \cite{pavlo} authors compare MR framework with parallel databases and notice that MR is compute intensive and high task startup costs dominate the execution time of short duration jobs. In \cite{floratou2012}, authors compare Hive with SQL Server PDW and observe that although Hive achieves better scalability, high CPU overhead associated with the RCFile format in Hive results in a slower query performance as compared to the SQL Server PDW. In \cite{floratou2014}, authors compare Hive and Impala and attribute the disk I/O efficiency, long running daemons and run time code generation in Impala as the reasons for its better performance than Hive. In \cite{icpe} authors compare Shark, Hive and Impala and observe that Impala exhibits the best CPU efficiency and the join performance worsens as the cluster size increases due to the increased data shuffle. In \cite{bigfun}, authors propose a social media inspired micro benchmark to compare AsterixDB \cite{asterix}, System-X, Hive-MR and MongoDB systems. Experimental results show that MongoDB becomes unstable for large aggregations due to memory issues and the lack of index support in Hive-MR causes point and range queries to be expensive. In \cite{titans}, authors evaluate MR and Spark for iterative and batch workloads using micro benchmarks to show that the CPU overhead associated with the de/serialization of intermediate results is the primary resource bottleneck.

To the best of our knowledge, this is the first work to thoroughly evaluate, understand and compare the performance of Drill, Phoenix and Spark SQL v2.0 systems for SQL workloads using standard analytics benchmarks including WDA and TPC-H. In addition, we compare the performance of aforementioned systems with Impala, a mature and well studied \cite{floratou2014, icpe} SQL-on-Hadoop system. 
\section{Discussion and Conclusion}
In this section we summarize the strengths and weaknesses of evaluated systems and the lessons learned from this study.
 
\textsc{\textbf{query optimizer.}} The query optimizers in Impala and Spark SQL generate simple and efficient execution plans by evaluating and selecting from a variety of join strategies including semi-join, anti-join, etc. However, we note that the cardinality estimates can be significantly off in Impala, resulting in expensive join order selection in some cases. The cost-based query optimization is still in its nascent stages in Phoenix; hence, users need to: 1) define the join evaluation order, and 2) choose the join algorithm to be used. The query optimizer in Drill can generate complex and inefficient execution plans, especially for the correlated sub-queries.

\textsc{\textbf{query execution engine.}} The Impala execution engine has the most efficient and stable disk I/O sub-system among all evaluated systems, as demonstrated by the lowest scan operator times for both storage formats. Although the Drill execution engine (with the columnar data model for in-memory processing) is optimized for the on-disk columnar data storage format (parquet), the scan operator is the principal contributor to the query RT in the parquet format. In addition, the scan operator becomes a performance bottleneck in Drill for the data stored in the text format with an unstable behavior characterized by the high scan operator wait times. In comparison with other evaluated systems, Phoenix has a notably larger data footprint due to the HBase HFile storage format, resulting in expensive full table scans.

The join and aggregation operator implementations in the Drill query engine harness all available CPU cores and achieve the shortest processing times among all evaluated systems. In contrast, the use of a single CPU core to perform the join and aggregation operations in Impala, results in sub-optimal resource utilization; however, ongoing efforts to enable multi-core execution for all operators \cite{impala-jira} should lead to performance improvement. Joins are costly in Spark SQL due to the choice of SMJ as the primary join operator, which requires an expensive sort operation prior to the join operation.

The data-exchange operator contributes nominally to the query RT in Impala, Spark SQL, and Drill. However, the client coordinated data-exchange operation is the primary performance bottleneck in Phoenix, making it ill-suited for join-heavy workloads that shuffle large amounts of data.  

The results from this study can be utilized in two ways: (i) to assist practitioners choose a SQL-on-Hadoop system based on their workloads and SLA requirements, and (ii) to provide the data architects more insight into the performance impacts of evaluated SQL-on-Hadoop systems.
{\bibliographystyle{IEEEtran}
\bibliography{IEEEabrv,comparator}
}
\end{document}